\documentclass[acmtr]{acmtrans2m}	

\usepackage[dvipdfm]{graphicx}

\acmVolume{1}
\acmNumber{1}
\acmYear{13}
\acmMonth{January}

\newcommand{\BibTeX}{{\rm B\kern-.05em{\sc i\kern-.025em b}\kern-.08em
    T\kern-.1667em\lower.7ex\hbox{E}\kern-.125emX}}

\title{A Computational Model\\for the Direct Execution of General Specifications\\with Multi-way Constraints}
\author{TOSHIO FUKUI\\Metalogic, Inc.}

\begin{abstract}
In this paper, we propose a computational model for the direct execution of general specifications with multi-way constraints.
Although this computational model has a similar structure to conventional constraint programming models, it is not meant for solving constraint satisfaction problems but rather for the simulation of social systems and to continue to execute assigned processes.
Because of this similar structure, it is applicable to the spectrum of the constraint solver, which is purple in this model.

Our computational model is not a declarative programming tool, unlike conventional constraint programs.
However, it can correspond to human logic when designing several applications or to design methods such as the data-oriented approach applied in many fields, including business systems, graphical user interfaces, web services, simulation, and data processing.
Essentially, it is a technology that can speed up the construction of large-scale network systems.

This model can be efficiently executed to directly describe design content in a simple way.
\end{abstract}

\category{D.2.0}{Software Engineering}{General}

\category{D.3.2}{Programming Languages}{Language Classifications}

\category{D.3.3}{Programming Languages}{Language Constructs and Features}

\category{F.1.1}{Computation by Abstract Devices}{Models of Computation}

\category{F.1.2}{Computation by Abstract Devices}{Modes of Computation}

\category{F.4.1}{Mathematical Logic and Formal Languages}{Mathematical Logic}

\category{I.2.2}{Artificial Intelligence}{Automatic Programming}

\terms{Theory, Languages, Design, Algorithms, Performance}

\keywords{automatic programming, computational model,
computational theory, constraint programming,
language design}

\begin{document}
\setcounter{page}{1}

\begin{bottomstuff}
Author's address: Metalogic, Inc.,
26-14, Toyotamaminami 3-chome, Nerima-ku, Tokyo, Japan;
email: fukui@metalogic.co.jp.
\end{bottomstuff}
\maketitle

\section{Introduction}

A great deal of research is performed on software under the banner of Computer Science.
Every possibility of computational models \cite{prgth1,prgth2,cstrpapers,cstrlogic,cstr}
has been researched.
However, the basic problem, which cannot be resolved by coding work, is handled through personal experience and ideas.
This problem, which is yet impossible to program mechanically,
is generally resolved by the components of the software and is progressing with great precision, but we cannot currently say that there is any basic solution for this.

Here, instead of conventional coding work, we advocate a programming paradigm in which operations are performed by directly describing the requirement specifications.

In this document, the development target is the business system and the program assigned to the computer, i.e., the business details, is the business flow.

\section{Installation}

\begin{figure}[tb]
\begin{center}
\includegraphics[keepaspectratio, scale=0.75]{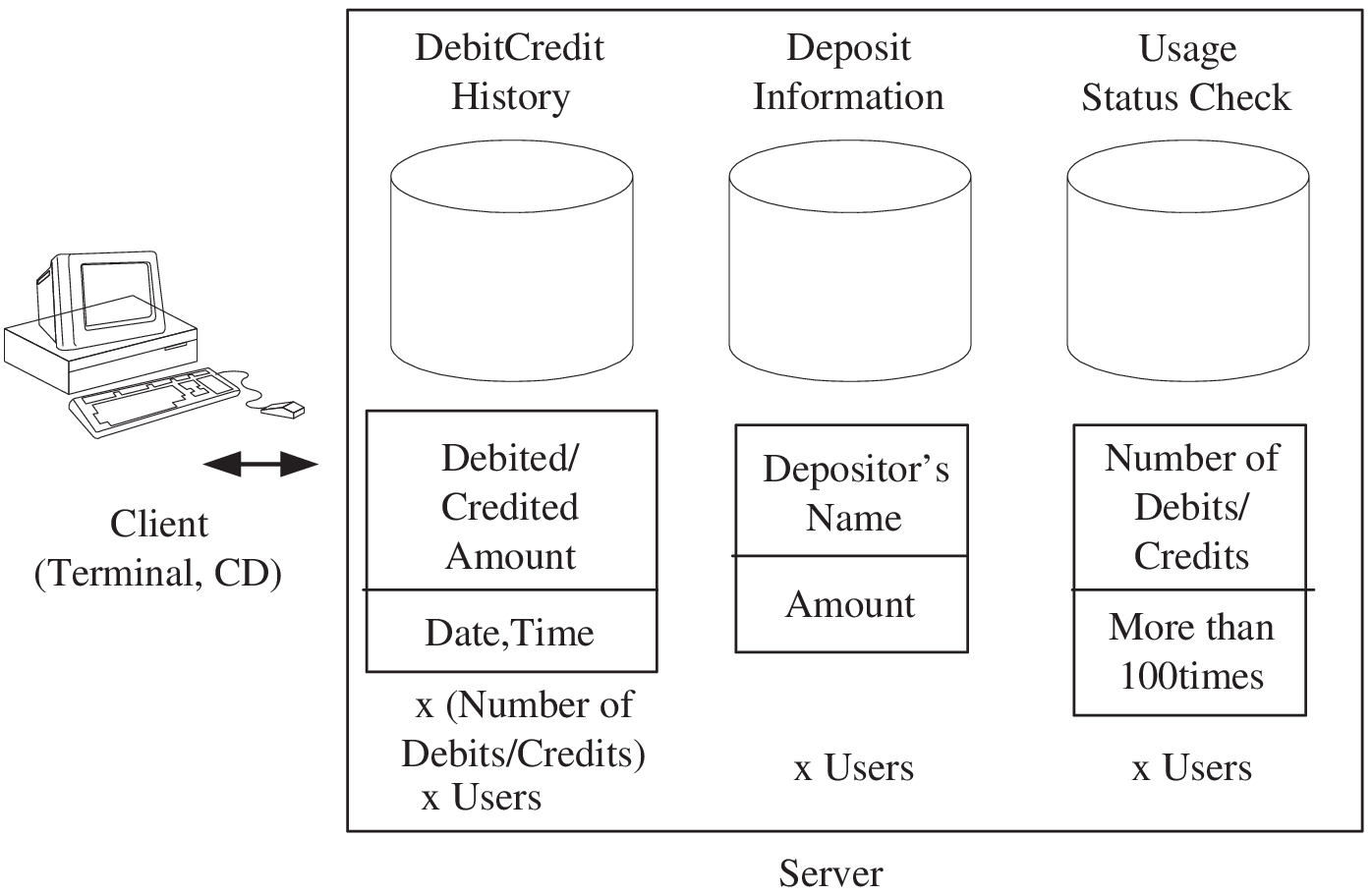}
\end{center}
\caption{Simplified Banking System}
\label{simple_bank}
\end{figure}

First, we describe the design work and operation on the computer using examples.
For instance, we will consider the simplified banking system shown in Figure \ref{simple_bank}.
The client-side shown in the figure can also be handled collectively by this method,
but the subject will be pursued by targeting only the server-side to simplify the explanation.

Generally, a technique based on data-oriented approach \cite{design2,design1} is used while designing, and a mechanical design procedure forms the data structure and data processing program from the specifications.
However, some homespun coding work is then performed to realize the data processing details.
If we put the permanent problems aside, a classic method for this coding work is to manage deposit information, debit and credit histories and check the usage status by allocating these operations to their respective structures or objects and providing the interface required for the operations.
Here, for the specific action to describe each operation, the CPU operation is said to be programmed from personal experience and ideas.

This technique is based on mainstream procedural programming, but there are different techniques using declarative programming.
Declarative programming enables us to relegate the actual resolving to the computer by describing the data relation.
However, in declarative constraint programming \cite{cstrpapers,cstrlogic,cstr}, large multivariate simultaneous equations have to be solved by the computer.
This presents the problem of execution efficiency and the complexity of execution maxims, and consequently the banking system was not developed on such a method.

Constraint programming overcomes this problem using mathematical algebra.
It is difficult to know whether the inside of a business system can be entirely described by expressions of an algebraic character because the intended purpose is to solve constraint satisfaction problems.
For instance, the algebraic approach cannot use processing where an immediate reverse-computation exists without a calculation to accompany the loop and complex conditional branching.
Thus, the coding problem is to achieve the present programming level of execution efficiency with this technique, which can allow a wide description.

As an example of an efficient constraint programming solver, we know of local propagation
\cite{deltablue,incsolver,skyblue,deltaup,hier},
which is classified as a blue algorithm,
and its application to user interfaces has been widely researched.

If we reconsider this established subject and review the design process, the following method is natural.
First, we shall define the data structure.

\begin{quote}
\begin{tabbing}
\textbf{struct} \{ \\
\hspace{1cm} \=	string $DepositorsName$; \\
	\> int $Amount$; \\
\} $DepositInformation[]$;
\end{tabbing}
\end{quote}

\begin{quote}
\begin{tabbing}
\textbf{struct} \{ \\
\hspace{1cm} \=	int $DebCreAmount$ = 0; \\
	\> string $Date$ = currentdate(); \\
	\> string $Time$ = currenttime(); \\
\} $DebCreHistory$[$Userid$][];
\end{tabbing}
\end{quote}

\begin{quote}
\begin{tabbing}
\textbf{struct} \{ \\
\hspace{1cm} \=	int $NumberOfDebCre$;\\
	\> bool $Morethan100times$;\\
\} $UsageStatusCheck$[];
\end{tabbing}
\end{quote}

Here, the amount will be a credit when positive and a debit when negative.
The relation between the deposit information and debit/credit history can be described by a formula.

If this formula is called the constraint, it can be written as follows.

\begin{quote}
$DepositInformation[j].Amount =$
\[
\sum_{i=0}^{vsize(DebCreHistory[j])} DebCreHistory[j][i].DebCreAmount
\]
\end{quote}

Here, $vsize()$ is the function for obtaining the array size and variable $j$ can take any value.
The relation between the deposit information and debit/credit history can be described by a similar formula because $<$number of times$>$ $=$ $<$ size of debit/credit history$>$.
Thus:

\begin{quote}
$UsageStatusCheck[j].NumberOfDebCre = vsize(DebCreHistory[j])$\\
$UsageStatusCheck[j].Morethan100times =$
\begin{quote}
$UsageStatusCheck[j].NumberOfDebCre > 100$
\end{quote}
\end{quote}

Seeing that this method is similar to the conventional local propagation, the computation will proceed after the computer determines the details of the given formula or the computation sequence.
However, the procedure for processing the formula here depends on the specifications, business flow, and business details given by the customer.
That is to say, these constraints already contain the actual data computation procedure, so there is no need to consider this in our method.

Although this is similar to constraint programming,
the decision of the constraint solution route at the problem solving stage is not needed,
and the constraint satisfaction problem is not solved.

Further, if the computational procedure is the business flow itself, and the computer understands the details sufficiently to change the procedure, it means that the details of the business that will be carried out by the human operators can be automatically analyzed.
Except in special cases, this is not possible with the current technical level.

Therefore, it is possible to arrive at the necessity of the existence
of a multiway computation mechanism by using a constraint equation
without trial and error in order to resolve 
non-constraint satisfaction problems whose purpose is different from that of
conventional constraint programming, including non-algebraic expressions.
This mechanism is indispensable to a blue algorithm
\cite{deltablue,incsolver,skyblue,deltaup,hier},
which is the most basic and efficient form in the constrain solver;
it is incapable of selecting the calculation route on
the basis of the calculation results and of preprocessing the planning.

Simply speaking, the method given in this example may have a two-way or one-way computation mechanism, but here we need to associate it to the general banking system.
We are able to select from several databases, and the selection may be done by the customers.
Depending on the case, the data may be registered in another database if the invalid access count reaches 5.
Therefore, depending on the data content, a dynamic direction must be selected at the time of execution and a multi-way computation method is required for this constraint solution.
However, the direction decision at execution can only be made beforehand by the designer.

In this case, for the above constraints $C$, if all variables $V$ are considered as $V_{all}$,
the constraint solution will be executed only if the constraints $C$ are related to variables
$V_c \subset V_{all}$ and the updated variables $T \subset V_c$ are generated.
This multi-way computation can be executed efficiently without fail by simply executing the operation to set the least related $V_{cn} \subset T$ variable to satisfy each constraint $C_n$.

Optional variable $j$ in this simplified banking system is the index of the arrays and is a single constraint problem.
However, if the variables $V$ are changed and become variables $T$, there is no need for any special computing to get the values to match the changed details.

In documents \cite{deltablue,incsolver} on the incremental blue algorithm and its later, improved form \cite{skyblue,deltaup},
a {\it walkabout strength} that uses a {\it locally-predicate-better} comparator is employed, and during the planning, methods such as a back tracking search of all the constraints $C$ are performed.
Even document \cite{hier} has planning and execution phases for local propagation using the comparator, though on comparison it behaves in the same way as our method, and so the validity of the problem or acyclic route detection becomes a major challenge.
It is prerequisite that the structure of the constraints of a targeted problem different from this method is unclear, and ultimately the unknown equation has to be solved, which makes the solution complex.

However, the comparator cannot be implemented for the problems targeted in this method, and the constraint solution route is already provided.
This means that there is no need for a mechanism to decide the solution route, as is commonly present in the red-orange-yellow-green-blue constraint solver spectrum, and only the pre-determined procedure is executed.

When a certain event occurs, there are $n$ pieces of the expression necessary, and if the time necessary to solve one expression is taken as the unit time, the constraint solver executes the shortest expression in $O(n)$ time units.
This is different to the solvable problem in an conventional constraint solver, and it does not solve the constraint satisfaction problems in the conventional sense.
Moreover, determination of the method route becomes unnecessary because it is possible to include a special non-algebraic function to freely describe the programming.
Thus, it is not necessary to interpret the meaning of the expression in this function.

Therefore, because the constraint satisfaction problem is not solved, it is difficult to say whether to enter the category of constraint programming.
However, due to the structural resemblance, our method can be classified as a constraint solver, and the distinguishing part resembles the structure of the blue part.
Here, a new color was required to be installed to suit the spectrum of the constraint solver, and it is purple.
However, only the structural part is similar in the end, and we do not achieve the solving of the constraint satisfaction problem.

To simplify the solution, in the above banking example,
a solution is required to generate a simple formula for such processing.
The program actually used in the business system does not involve any complex computation,
but merely transforms the data and repeats the transfer process to simulate a part of the social structure with the aim of enabling us to reference the simulated result status.
Therefore, we can presume that in the actual system, several conditional judgments and diversions are used in multi-way computation formulas,
so this simplified banking example is not special.

A summary of the above characteristics is as follows:
\begin{itemize}
\item In the given constraints, the procedure for processing the solution is considered and there is no contradiction.
\item The constraints here need not be algebraic, and contain the function that does the programming.
\item The decision or judgments of the computation sequence, which are mandatory in conventional multi-way computation, are neither needed nor can they be implemented.
\end{itemize}

This is a design method in itself, and it is difficult to call it declarative because the processing procedure of the solution is given by the user and has the main characteristics of constraint programming.
That is why, from here on, we call this method non-declarative constraint programming, or a computation model with a purple constraint solver.

The proposed method only handles problems whose solutions are clear.
Consequently, the nature of problems handled in multi-way computation according to the constraints in this method is essentially different from declarative constraint programming, except for the fact that both handle constraints.
However, it is certain that the programming field where this method is applicable exists, and the previously mentioned banking system is a typical business system; therefore, this method can directly correspond with the processing sequence specifications given to the computer at design-time.
Data processing and reshaping are the main processes in the business system, which means that the problem will not be solved by computing an equation.
Most of the software development work consists of system development in general, and other developments are just a part of the development volume, a fact learnt from hands-on experience on the development site.

Actual system development consists of many complex variables, and when specific data is changed, the data flows while transforming through multiple linked constraints.

\begin{figure}[tb]
\begin{center}
\includegraphics[keepaspectratio, scale=0.45]{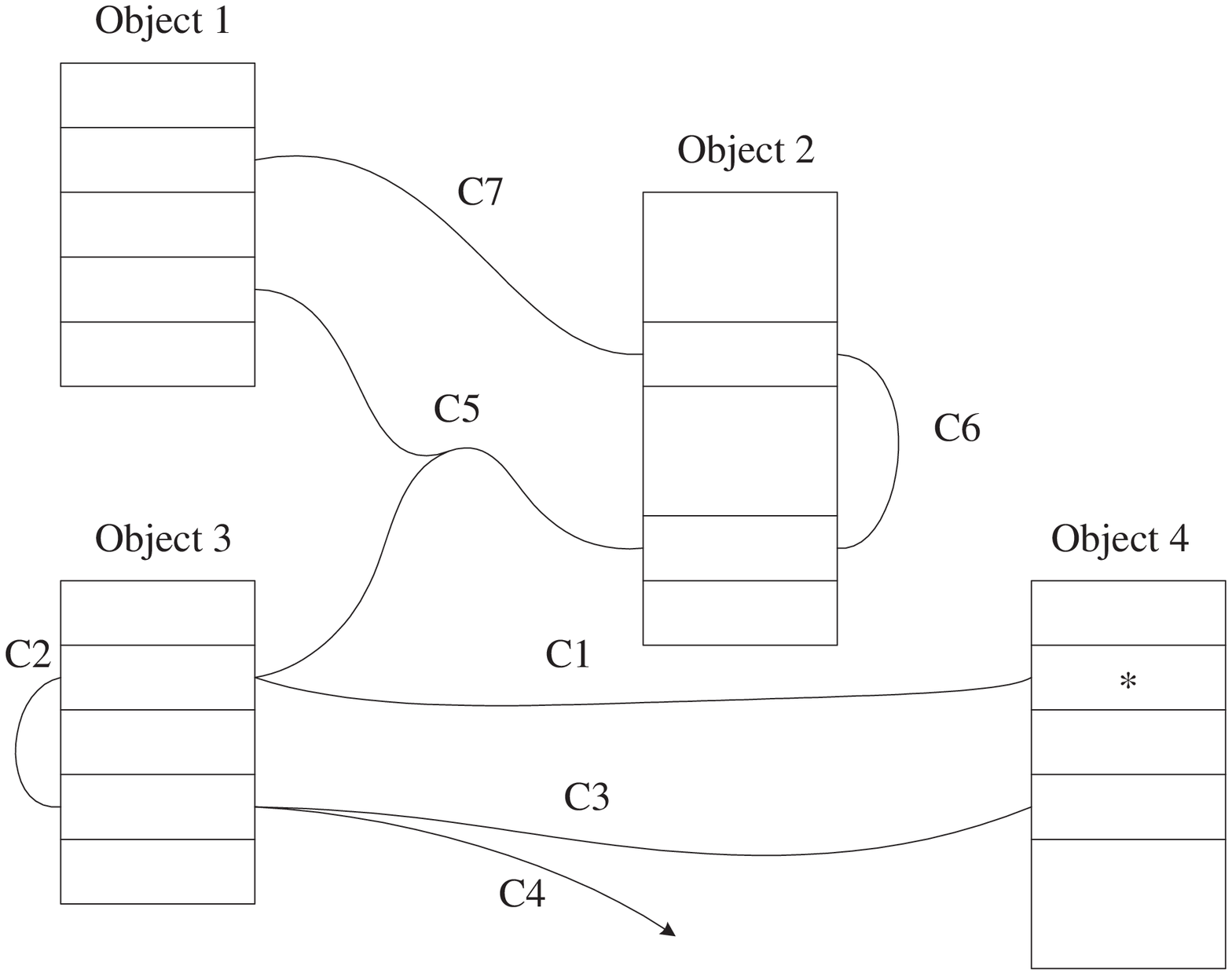}
\end{center}
\caption{An example of the complex system}
\label{complex_system}
\end{figure}

We clarify the meaning of multi-way computation by presenting an outline of a complex system in Figure \ref{complex_system}.
Lines C1 to C7 are the constraints joining the variables, and there is no limit to the number of variables that can be referenced by one constraint.
A wide description capability for changes in the substitution destination by the condition comparison is required from the multivariate constraints in the system description.
Here, when the * marked value in Object 4 is changed, the constraints are executed in the order C1 to C7
if the parallelism of the divisions is ignored and they are executed in a proper sequence.
That is, the constraint solution sequence, which is the sequence of computation in this method,
is not limited and is multi-way, and this sequence is determined from the connected constraints and the changed data.
Here, this solution sequence is called the data flow.
Its multi-way nature is a general feature of constraint programming, and as the user considers the data flow while entering the constraints, the process for deciding the route is not required and no other process besides execution of the given computation is carried out.

The termination problem does not exist in our proposed model,
unlike conventional constraint programming\cite{termproblemcstr}
and generic programming languages\cite{termproblemgpl},
because our method does not generate a new execution flow,
and it continues to execute only the given data flow.
Because the proposed design is a business flow,
it excludes the case in which, when it goes is wrong,
it executes and terminates appropriately.
The validity of the behavioral specification of the application corresponds to the validity of execution.

Our programming language is executed directly as design information is input as a program, and the compilation doesn't exist.

Other similar computation models besides constraint programming include one-way or two-way computation models, such as programming methods based on spread sheets \cite{spreadsheet} or data flow machines \cite{dataflow}.
In document \cite{spreadsheet}, multi-way constraints are handled,
but multi-way constraints that move freely without any limitations, as given here,
cannot be treated.
In addition, the computation model described in document \cite{spreadsheet} does not compute continuously from the linkage analysis in a series.
This is because of the difference in the adaptive target.

Besides, the document \cite{alchemy} is a research near this.
The document \cite{alchemy} has treated the Alloy specification of the relation base, and the compiler generates the updating operation of the data base that CPU does.
The difference with our programming language is the following two.
The input program doesn't satisfy the mathematical relation, and contain the procedural programming language functions.
The input program is executed directly without the conversion works.

There is Comet\cite{cstrpapers} that Professor Pascal Van Hentenryck developed as a programming language that builds in the constraint solver.
Because this has aimed to solve the constraint satisfaction problems as well as a blue algorithm, it is fundamentally different from our method.

\section{Software Categorization for Applicable Targets}

We need to categorize software to prepare the applicable targets of this method.
In document \cite{spreadsheet}, software was categorized into interactive and batch applications,
and interactive applications were the target field.

Although batch applications will be the main target of our purple constraint solver, there are some areas in which this method is not suitable.
For this reason, we need some other scale to categorize applications.

\begin{figure}[tb]
\begin{center}
\includegraphics[keepaspectratio, scale=0.55]{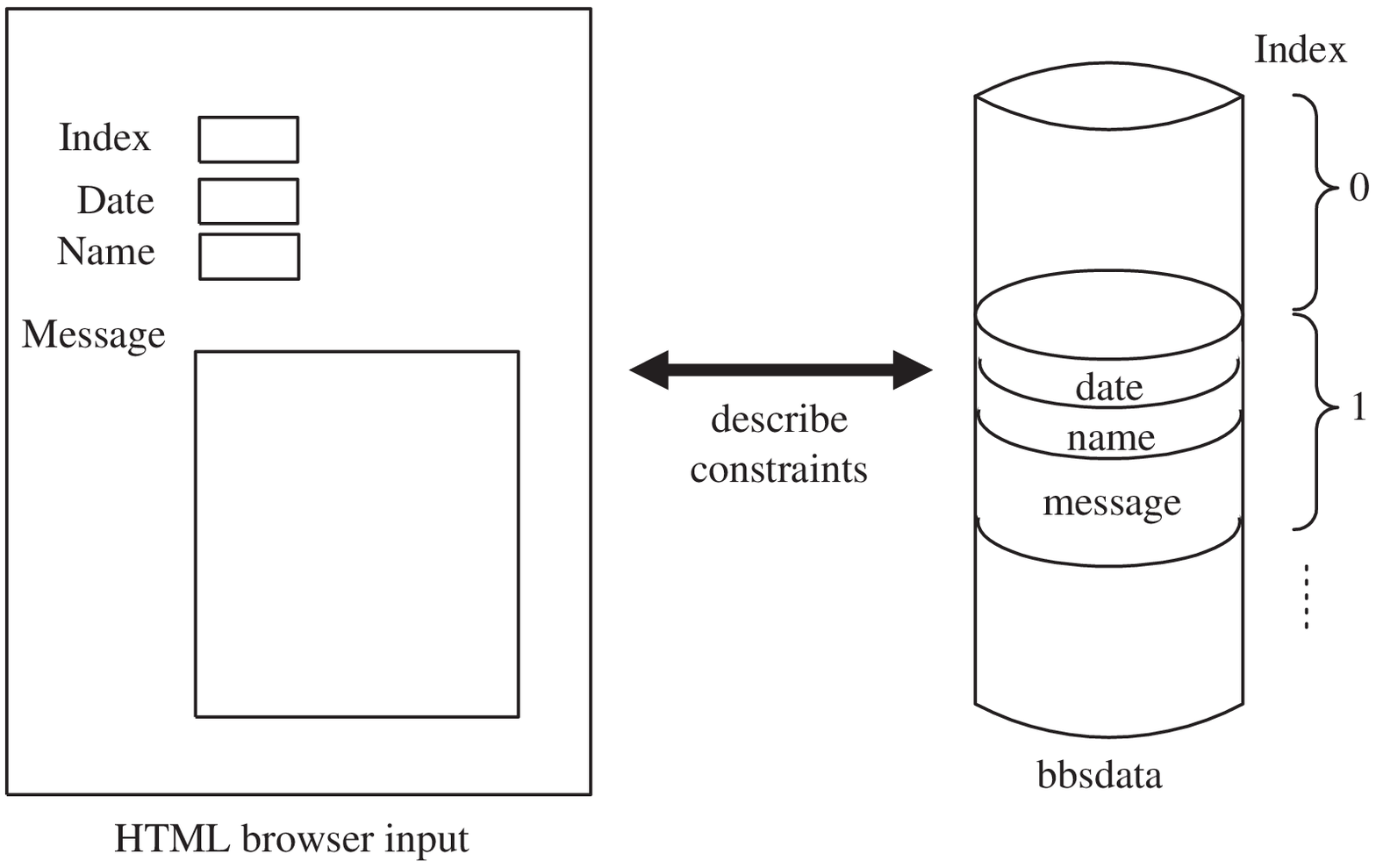}
\end{center}
\caption{Input and Stored Data of Web Application}
\label{web_relation}
\end{figure}

\begin{figure}[tb]
\begin{center}
\includegraphics[keepaspectratio, scale=0.55]{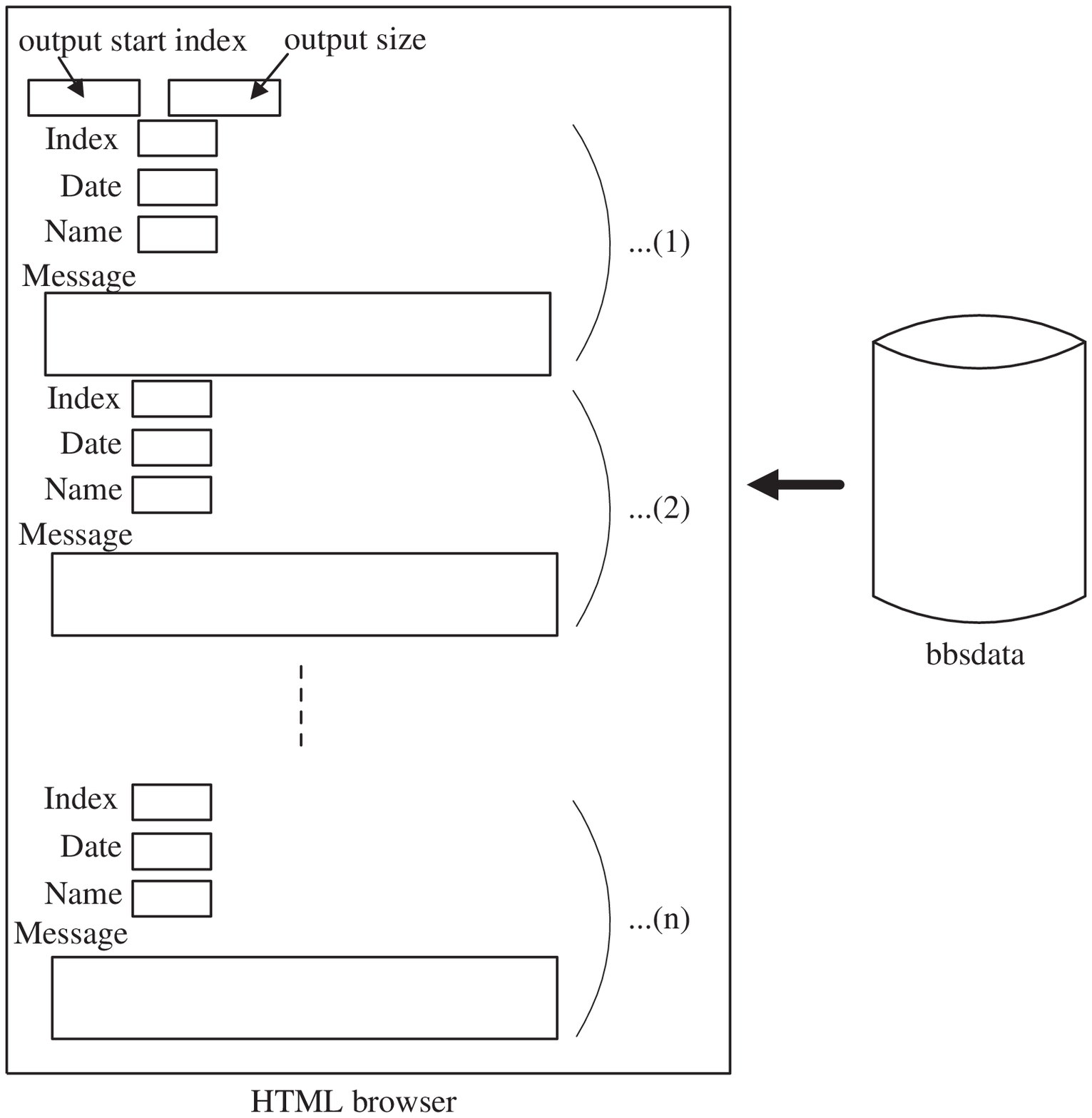}
\end{center}
\caption{HTML Output of Web Application}
\label{web_output}
\end{figure}

The input/output structure of a simplified bulletin board-format web application is shown in
Figure \ref{web_relation} and Figure \ref{web_output}.
For input and output of an HTML system, the HTML can be assumed to give layout information and various properties of the parts can be presumed as variables.

If we focus on the relation between the properties and the corresponding actual data variables, we can see that it has the same structure as the banking system.
This is not only true in HTML: the structure is generally the same in a graphical user interface (GUI) and printing for document output is also similar.
The layout or the GUI operation itself can be separated from the framework, so this method can use a wide range of user interfaces.
If the HTTP protocol is handled as it is, the data and display cannot be handled two-way and HTML does not have any problem, essentially because it can be separated into the data and layout.
In \cite{spreadsheet}, the construction of a complex GUI is considered as a prerequisite and can be considered as a special case of this method.
However, as the user interface in a business system has a simple document base, it is much simpler.

Generally, a business system consisting of a web system and a client server is made by integrating them through a network.
This interface including a network can be abstracted as variables in the same way as a user interface, so all the elements can be handled by this method for system development.

Consequently, our method is suitable in the following fields.

\begin{enumerate}
\item Business systems
\item User interfaces (GUIs)
\item Web systems, service programs
\end{enumerate}

On the other hand, this system is not suitable for some other applications.
These are mainly computation processes for numerical computing.
Computation processes include video and audio processing.

Here, the characteristics of an application are simply classified, and the first is called
``Data operational applications'' while the latter is called ``Computational applications''.
Data operational applications do not have many computation processes, but their core operations are the four arithmetical operations, data transforming, and copying.
As a result, in many cases, the relation between data can be simply described as constraints.

If we discuss further classifications, we can consider applications like editors or word processors as computational applications.
However, for example, many computational applications can be incorporated as parts of data operational applications.

Then, how should we include computation of the total in the accounting computation problem shown in Table \ref{accounter_base}.

\begin{table}[tb]
\begin{center}
\caption{Accounting Computation}
\label{accounter_base}
\begin{tabular}{|r|r|r|c}			\cline{1-3}
Price	& Number	& Sub Total	\\	\cline{1-3}
100	& 2	& 200	& $\cdots [1]$	\\	\cline{1-3}
200	& 3	& 600	& $\cdots [2]$	\\	\cline{1-3}
300	& 2	& 600	& $\cdots [3]$	\\	\cline{1-3}
50	& 3	& 150	& $\cdots [4]$	\\	\cline{1-3}
\multicolumn{2}{|c}{Total}
		& \multicolumn{1}{|r|}{1550}
			&		\\	\cline{1-3}
\end{tabular}
\end{center}
\end{table}

\begin{figure}[tb]
\begin{center}
$total = \sum_{i=1}^4 subtotal[i],$ \\
$subtotal[i] = price[i] \times number[i],$ \\
$price[1] = 100, price[2] = 200,$ \\
$price[3] = 300, price[4] = 50,$ \\
$number[1] = 2, number[2] = 3,$ \\
$number[3] = 2, number[4] = 3$
\end{center}
\caption{Constraints of Accounting Computation}
\label{accounter_by_cstr}
\end{figure}

If the solution is declarative in constraint programming, with Price, Number, and SubTotal as the respective array variables and indexes from $[1]$ to $[4]$, the problem can be described by the constraints shown in Figure \ref{accounter_by_cstr}.
This is a typical computational application where the total is obtained from the constraint solver.

However, if the data processing is carried out in the actual business in the same way as in the simplified banking system, the data will be stored in a database and, as the data is input continuously, the computation result will be continuously updated.
This problem needs to be solved continuously, and so differential computation is enabled.
This is a common characteristic of business systems and is common in most of the data operational applications described here.

As a result, by linking the respective constraints $C$ to the variables $V$, an approach depending on the variation of variables is possible, and in this case we can regard it as a data operational application.

\section{Description of Constraints}

Although a mechanism for determining the solving method route is not necessary in this model, it is necessary to prepare a solver for each of the constraints.
If we consider the constraint to be an equality, the relation between the left side $a$ and the right side $b$ can be described as:
\[
	a = b
\]
In executing it, if the left side $a$ changes, the right side $b$ must also change.
However, the way to determine the value of the right side and whether the values of the variables included in the formula $b$ can be determined or not depends on the structure of the constraints and the condition at the time of execution.
As $a$ and $b$ can be abstract data structures or objects, they are not simply single variables.

Taking a sales order system as an example,
if inventory $a$ decreases, this could imply that order list $b$ has increased.
However, if the ordered product is cancelled, the order list $b$ decreases and inventory $a$ also returns to its original value.
This kind of two-way computation is often required when solving constraints, but the computation methods are not necessarily simple.

The basis of this computational model lies in setting the constraints so that no inverse solution method takes place in solving problems.
Therefore, although it is not necessary to define inverse operations except when required, there are many cases where a solution method exists even though the problems cannot be handled mathematically.
The constraints that do not have an inverse operation defined are considered to be one-way constraints.

The constraint condition with an implicit constraint solution method $y = f(x)$,
for which the system can automatically prepare the constraint solution method, is described as:
\[
	[ \; y = f(x) \; ]
\]
and the constraint condition with an explicit constraint solution method $y = f(x)$,
for which the programmer indicates the constraint solution method, is described as:
\[
	[ \; y = f(x) \; ] \neg\Rightarrow [ \; y \leftarrow f(x); \; x \leftarrow f^{-1}(y); \; ]
\]
As a result of this introduction, it becomes unnecessary for the left and right sides of the constraint to have any mathematical meaning.
Although only a single output is assumed here while solving the solution formula, there is no such limitation and there can be multiple outputs.

In general, methods based on unification, the Knuth-Bendix completion procedure,
or the usage of the constraint solver has been used to guarantee the solution route in declarative programming.
Therefore, the expression and its elements had to have a mathematical structure in order for these to be applied.

In non-declarative constraint programming, the expression is only sequentially simply solved,
because we only know that the solving route and the termination are guaranteed.
Therefore, because the constraint can be treated by simple processing of the evaluation and the expression transformation,
the description need not be limited.

The functions $f()$and $f^{-1}()$ build the procedural programming description because they are not algebraic.
Therefore, it is possible to build an efficient description of the procedural programming language including the loop and the conditional branching of the constraints.

This makes it look as if describing the behavior of CPUs is indispensable, just like the procedural programming language,
but in fact, the essence of the overall program is well organized by the constraints as a result of localizing the procedural programming.
The reuse rate is also high because it is not a modulation of the business but the data transformation, and an excellent modulation environment is being offered compared with conventional procedural programming.

In the case of the implicit constraint method, the solution method is the solution of the equation.
In this case, it is taken for granted that complex mathematical or algebraic solution methods need to be prepared, but as the tendency of general programming on many business systems is towards accounting computations, simple substitutions, easy addition, subtraction, and string operations rather than mathematical problems, formula transformation by means of simple transposition handling is sufficient.

Though it is also possible to handle an increasing number of variables in multi-way computation, as the data destination is mostly selected according to the computation result in business systems, we introduce if-then-else decisions in the above computation formula to divert the conditions.

In this model, the necessary computations are achieved as a whole by combining the one-way, two-way, and multi-way constraints in a complex manner.
As this system applies mainly for business systems that achieve social structure simulations, the structure becomes very complex with various intertwined constraints.
However, as the complex structure corresponds to the specifications of the business, nothing extra has to be considered at the time of describing the constraints.

Although operators or functions are necessary for matching and array operations in actual constraint descriptions, they need not necessarily be new and can be included above in $f(x)$.

\section{Purple Constraint Solver}

The above concept has been reconsidered and generalized.
The behavior of most of the data operational applications, such as business systems, can be described as follows.

\begin{enumerate}
\item All the data on the computer is in the form of solved problems and the computer need not work as long as there are no changes to the data.
\item Appropriate processing is done when there is a change to the data and the problem is solved.
\item Returns to (1).
\end{enumerate}

If we express this with constraints $C$ and variables $V$, then (1) to (3) can be rewritten as follows.

\begin{enumerate}
\item All the constraints $C$ are fulfilled and the computing machine has stopped until changes are added to the related variables $V_c \subset V_{all}$.
\item Constraint solving is executed when changed variables $T \subset V_c$ are generated and the value of variables $V_{cn} \subset T$ are adjusted until the various constraints $C_n$ are satisfied.
\item Returns to (1).
\end{enumerate}

\begin{figure}[tb]
$solver$( $V$, $T$, $C$ ) \{\\
$S \leftarrow T$\\
\textbf{while} $S \neq \emptyset$ \{\hspace{1cm}
\begin{quote}
	$x \leftarrow \exists s \in S$ $\;\;\;\cdots$ (1)\\
	$Q \leftarrow \forall c \in C$ has
\begin{quote}
		some dependent values related to $x$
\end{quote}
	\textbf{while} $Q \neq \emptyset$ \{
\begin{quote}
		$\exists c \in Q$ is taken away to solve,
\begin{quote}
		$\forall v \in V$ is adjusted to satisfy $c$. $\;\cdots$ (2)
\end{quote}
		$S \leftarrow S \; \cup$ adjusted $\forall v$
\end{quote}
	\}\}\}
\end{quote}
\caption{Algorithms for the Purple Constraint Solver}
\label{main_algorithm}
\end{figure}

This is handled as an assumed operation.

Here, the constraint $C$ is assumed as follows.
\[
[ \; condition \; ] \neg\Rightarrow [ \; solver1; \; solver2; \; \ldots \; solvern; \; ]
\]
One executable $solver$ is executed on approval of the evaluation result of $condition$,
i.e., $condition \longrightarrow \neg true$.
However, the existence of the executable $solver$ is a precondition.
There are no limitations, as this condition satisfies the given constraint.

Furthermore, if we define the solver by considering that the data flow is assigned,
the problem will be merely to solve the constraint $C_n$ related to variable $V_{cn}$ unconditionally without considering the order.
Thus, it becomes a simple algorithm as shown in Figure \ref{main_algorithm}, where the overall variables $V_{all}$ are $V$, all the variables that change in value are $T \subset V_c$, and all the constraints are $C$.
However, $S$ and $Q$ are sets.

It is the user's responsibility to set the constraints so that they can be solved as shown in Figure \ref{main_algorithm},
and so if the algorithm becomes unexecutable in the middle it will be due to a programming error.
However, as the operation order and constraint solving process are taken into account automatically by complete compliance in the design, constraints that are actually assigned will not end with contradictory or inefficient operations as long as there is no error.

The following three points should be noted.
\begin{itemize}
\item The constraints used here are simple addition and subtraction constraints, such as those in the first simplified banking system, but the data flow is very important in many programs representative of actual business systems and is made of many such addition and subtraction operations.
Therefore, in many cases, the constraints are simple operations of formula transformation and the methods for solving them are obtained for specific variables, allowing the two-way or multi-way computations to be freely advanced.
\item The operations for the elements in each set of the algorithm shown in Figure \ref{main_algorithm} can be processed concurrently, and each procedure in Figure \ref{main_algorithm} can also be executed in parallel.
This computational model essentially has perfect parallelism.
\item The complexity in this computational model is determined by the constraint solving procedure set by the user.
Therefore, the user can forecast and control the total complexity.
\end{itemize}

\section{Programming Language RIPPLE}

\subsection{Summary}

The programming language RIPPLE implements a mathematical formula data structure in the functional programming language LISP,
prepares a mechanism to convert the C-style grammar into S-expressions in the front-end, and integrates a programming language based on the algorithm in Figure \ref{main_algorithm} as its execution principle.

The reasons for selecting LISP in place of the conventional constraint programming language are as follows.
\begin{itemize}
\item A complex constraint solver, in the conventional sense, is not necessary in the computation model of this article,
and the only elements necessary for implementation are the symbolic processing for computing the formula.
\item The constraint-describing capacity increases by selecting a functional programming language; in fact, it becomes limitless.
\end{itemize}

RIPPLE executes the programming of the LISP language using the S-expression conversion mechanism through command prompt inputs.
Though registration of constraint conditions, functions, and function calls are performed by the execution of the LISP language, the execution of formula and constraint solution processing is performed through a function known as the exclusive formula processing mechanism.
The function creates a list of the variables whose constraint solution is required for internal execution.
Before displaying the ultimate output value, the constraint solution operation shown in Figure \ref{main_algorithm}
is executed until the convergence criteria are met.

The input grammar for RIPPLE, with the exception of the constraints, is shown in Figure \ref{grammar_main}.
This is simply an addition of the input constraints process in the C-language.
The RIPPLE-type system is the same as the Common LISP-type system,
because the mathematical formula data structure is used for internal implementations in this programming language system.
Therefore, in the external specification of RIPPLE,
only processing concerning the purple constraint solver is added to
the conventional programming language,
which is essentially the only new element.
Moreover, current RIPPLE doesn't implement the parallel processing.

\subsection{Constraints}

\begin{figure}[tb]
\begin{center}
\begin{tabbing}
\textit{execution-unit} $::=$				\\
\hspace{1.5cm} \= \textit{constraint-declaration}	\\
	\> \textit{generic-declaration}			\\
	\> \textit{evaluation}				\\
\\
\textit{generic-declaration} $::=$			\\
	\> C-language-style-function-declaration	\\
	\> C-language-style-variable-declaration	\\
\\
\textit{evaluation} $::=$				\\
	\> C-language-style-expression \texttt{;}	\\
\end{tabbing}
\end{center}
\caption{Main Grammar of RIPPLE}
\label{grammar_main}
\end{figure}

\begin{figure}[tb]
\begin{center}
\begin{tabbing}
\textit{constraint-declaration} $::=$				\\
\hspace{1.5cm} \= \texttt{[} \textit{condition} \texttt{] ;}	\\
	\> \texttt{[} \textit{condition} \texttt{] - > [}
		\textit{solvers} \texttt{] ;}			\\
\\
\textit{condition} $::=$					\\
	\> \textit{condition-expression}			\\
	\> \texttt{[} \textit{matching-variables}
		\texttt{]} \textit{condition-expression}	\\
\\
\textit{condition-expression} $::=$				\\
	\> C-language-style-expression \texttt{;}		\\
\\
\textit{solvers} $::=$						\\
	\> \textit{solver-expression}				\\
	\> \textit{solver-expression} \textit{solvers}		\\
\\
\textit{solver-expression} $::=$				\\
	\> C-language-style-expression \texttt{;}		\\
\end{tabbing}
\end{center}
\caption{Constraint Declaration Grammar of RIPPLE}
\label{grammar_constraints}
\end{figure}

The method of describing constraint declarations is shown in Figure \ref{grammar_constraints}.
The operators and the functions that can be used by constraints C-lang-style-expression do not have the limitation.
Our programming language is LISP based system.
Therefore, a different meaning is defined in the operator such as pointers, and the list data structure is expressible.
Details are omitted because it is not essential.

When the variable is changed,
RIPPLE evaluates {\it condition-expression} of the constraints related to the variable.
The variables of RIPPLE has information to the constraint that relates with the value.

When an evaluation result of {\it condition-expression} is not established for the constraint,
i.e., when the evaluation result is $false$, $null$, or $nil$, the {\it solver-expression} is selected and executed.
The selection at the time when an evaluation result is not established is always 1.
It is a bug that the execution of the solver doesn't satisfy conditional expression usually.
Therefore, after the solver is executed, it verifies it by revaluing conditional expression.

When {\it solver-expression} is one in the constraint definition, it is special.
When {\it solver-expression} is one, {\it solver-expression} need not be selected.
The solver need not understand the meaning of {\it condition-expression}.
Therefore, the programmer can use the complex user-defined function
including the branch and loop structure for {\it condition-expression}
and {\it solver-expression}.

When {\it solver-expression} is one,
the one-way constraint and the branch to the multi-way are described.
When there are two or more {\it solver-expression},
the simple bi-directional constraint and the multi-way constraint are described.
Because both can use it by combining,
the multi-way constraint with various structures can be unrestrictedly described.
Therefore, the constraint solver of RIPPLE has been achieved
by an extremely simple mechanism.

Any number of variables can be described in the {\it matching-variables}.
If a matching variable is included in evaluating a {\it condition-expression},
the matching variable is handled as an arbitrary value, and the value that could be matched
in evaluating the {\it condition-expression} is substituted.
The arbitrary value can be referred to as the same variable in the {\it solver-expression}.

The array index number, or the generalized index range, can be given as an example of the main value to be substituted in the matching variable.

\subsection{Abstract Function Value}

It is necessary to increase and decrease the array element in the description of the constraints.
For instance, the function $vsize()$ sets the processing method to simultaneously change the array size to the return value and specify the size of that array.
When substitution is directed to the return value of the function, the array size is changed.
The LISP part of RIPPLE can define the abstract function value in an arbitrary place.
The access function stored in the value by function $setf$ can be called.

The concept of an abstract function value is more complex, although it looks like the reference.
It simultaneously has the access function for substitution and the value as a data structure of the return value of the function.
When substituting it, the access function is called.
Here, because it is an abstracted value with the function, it is called the abstract function value.

\subsection{Sample 1: Simplest Computation Program}

The simplest two-way computation program on RIPPLE is shown below.

{\footnotesize
\begin{quote}
int $a = 5$; \\
int $b = 0$; \\
$[ \; a == b + 5; \; ];$
\end{quote}
}

\subsection{Sample 2: Simplest Multi-way Computation Program}

The multi-way computation program that executes the simplest conditional branching is shown.
The function $if()$ is the $if$ function of LISP.

{\footnotesize
\begin{quote}
int $a = 5$; \\
int $b = 0$, $c = 0$; \\
$[ \; if( \; a < 10, \; b == a, \; c == a ); \; ] \; -> \;
[ \; if( \; a < 10, \; b = a, \; c = a); \; ];$
\end{quote}
}

The variable at the substitution destination differs according to the value region,
and expresses the conditional branching.
The solver expression does not necessarily fill the conditional expression in this kind of constraint expression.
However, the purple constraint solver only executes the solver expression when the relating variable changes.
This is the operation originally intended.

\subsection{Sample 3: Accounting Computation Program}

The accounting computation programming
in Table \ref{accounter_base} is now described on RIPPLE,
as shown in Figure \ref{ripple_src_accounter}.
It is an example of the batch application in \cite{spreadsheet},
although this is a one-way computation.

\begin{figure}[tb]
\textbf{Program}

{\footnotesize
$dynamic$ $automatic$ \textbf{struct} $account\_struct$ \{
\begin{quote}
	int $value = 0$; \\
	int $n = 0$; \\
	int $ssum = 0$;
\end{quote}
\} $[0]$ $account\_data$; \\
\\
int $totalp = 0$, $total = 0$; \\
float $tax = 0.0$; \\
\\
int $account\_total$( array $x$ ) \\
\{
\begin{quote}
	int $i$, $n$; \\
	$n = 0$; \\
	\textbf{for} ($i = 0$; $i < vsize(x)$; $i$++)
\begin{quote}
		$n += x[i].ssum$;
\end{quote}
	\textbf{return} $n$;
\end{quote}
\}; \\

$[[i] \;\; account\_data[i].ssum == account\_data[i].value
 * account\_data[i].n; \; ];$ \\

$[ \; totalp == account\_total(account\_data); \; ];$ \\

$[ \; total == totalp * tax; \; ];$ \\
}

\textbf{Execution}

{\footnotesize
$tax = 1.05$; \\
$account\_data[0].value = 100$; \\
$account\_data[0].n = 2$; \\
$account\_data[1].value = 140$; \\
$account\_data[1].n = 3$; \\
$account\_data[2].value = 120$; \\
\hspace{2cm} \vdots \\
}
\caption{Accounting Program on RIPPLE}
\label{ripple_src_accounter}
\end{figure}

Some necessary explanations are given below.

\begin{itemize}
\item The initial values of the structure member are referred to when
 the array size is varied.
\item The qualifier $dynamic$ shows that the array size can be varied,
 and the qualifier $automatic$ shows that
 the array size is automatically adjusted by means of an index reference.
\end{itemize}

Insertion, and deletion are performed with array element operations.
For instance, the deletion operation
\[
account\_data[0..2] = account\_data[0,2..3]
\]
is possible, and the array is extended automatically by the insertion operation.

\subsection{Sample 4: Simplified Banking System}

The simplified banking system discussed earlier is now cast in RIPPLE,
as shown in Figure \ref{ripple_src_simplebank}.
Permanent data is omitted to simplify the problem.
Figure \ref{ripple_oper_simplebank} illustrates various operations.

\begin{figure}[tb]
{\footnotesize
$dynamic$ $automatic$ \textbf{struct} $deposit\_information\_struct$ \{
\begin{quote}
	string $depositors\_name =$ ``''; \\
	int amount = $0$;
\end{quote}
\} $[0]$ $deposit\_information$; \\
\\
$dynamic$ $automatic$ \textbf{struct} $debcre\_history\_struct$ \{
\begin{quote}
	int $debcre\_amount$ = 0; \\
	string $date =$ ``$2013/01/01$''; \\
	string $time =$ ``00:00:00'';
\end{quote}
\} $[1000][0]$ $debcre\_history$; \\
\\
$dynamic$ \textbf{struct} $usagestat\_chk\_struct$ \{
\begin{quote}
	int $number\_of\_debcre$ = 0; \\
	bool $morethan100times$;
\end{quote}
\} $[0]$ $usagestat\_chk$; \\
\\
int $debcre\_total$( array $x$ ) \\
\{
\begin{quote}
	int $i$, $n$; \\
	$n = 0$; \\
	\textbf{for} ($i = 0$; $i < vsize(x)$; $i$++)
\begin{quote}
		$n += x[i].debcre\_amount$;
\end{quote}
	\textbf{return} $n$;
\end{quote}
\}; \\

$[[j] \;\; deposit\_information[j].amount ==
debcre\_total( \; debcre\_history[j] \; ); \; ];$ \\

$[[j] \;\; usagestat\_chk[j].number\_of\_debcre ==
vsize( \; debcre\_history[j] \; ); \; ];$ \\

$[[j] \;\; usagestat\_chk[j].morethan100times == if($
\begin{quote}
	$usagestat\_chk[j].number\_of\_debcre > 100, t, null \; ); \; ];$
\end{quote}
}
\caption{Simplified Banking System on RIPPLE}
\label{ripple_src_simplebank}
\end{figure}

\begin{figure}[tb]
{\footnotesize
\textbf{User Register:}\\
$deposit\_information[0].depositors\_name = "User 1";$\\
$deposit\_information[1].depositors\_name = "User 2";$\\
$deposit\_information[2].depositors\_name = "User 3";$\\
\\
\textbf{Debit/Credit of User 1:}\\
debcre\_history[0][vsize(debcre\_history[0])].debcre\_amount = 10000;\\
debcre\_history[0][vsize(debcre\_history[0])].debcre\_amount = -5000;\\
\\
\textbf{Debit/Credit of User 2:}\\
debcre\_history[1][vsize(debcre\_history[1])].debcre\_amount = 20000;\\
debcre\_history[1][vsize(debcre\_history[1])].debcre\_amount = -10000;\\
}
\caption{Operations in the Simplified Banking System}
\label{ripple_oper_simplebank}
\end{figure}

\subsection{Sample 5: Simplified Banking System with Multi-way Computation}

To make Figure \ref{ripple_src_simplebank} practicable,
the storage data base is divided into two parts, which are selected according to user ID.
This example is used to show the essence of the system,
and is necessarily simplified in order to describe it.
Figure \ref{ripple_src_multiwaybank} shows the changes.

\begin{figure}[tb]
{\footnotesize
\textbf{struct} $debcre\_history\_struct$ \{
\begin{quote}
	int $debcre\_amount$ = 0; \\
	string $date =$ ``$2013/01/01$''; \\
	string $time =$ ``00:00:00'';
\end{quote}
\}; \\
\\
$dynamic$ $automatic$ \textbf{struct} $debcre\_history\_struct$
 $[1000][0]$ $debcre\_history1;$ \\
$dynamic$ $automatic$ \textbf{struct} $debcre\_history\_struct$
 $[1000][0]$ $debcre\_history2;$ \\
\\
$[[j] \;\; deposit\_information[j].amount == debcre\_total($
\begin{quote}
	$if( \; j<500, debcre\_history1[j], debcre\_history2[j] ) \; ); \; ];$
\end{quote}

$[[j] \;\; usagestat\_chk[j].number\_of\_debcre == vsize($
\begin{quote}
	$if( \; j<500, debcre\_history1[j], debcre\_history2[j] \; ) \; ); \; ];$
\end{quote}
}
\caption{Changes to Multi-way Banking System}
\label{ripple_src_multiwaybank}
\end{figure}

\subsection{Sample 6: Simple Web Application}

A simple bulletin board on the web is operated as a content example of a user interface on RIPPLE.
The structure of this application is shown
in Figure \ref{web_relation} and Figure \ref{web_output}.
This program is a one-way computation example, as there is no two-way protocol in HTTP.
In this case, the multi-way computation is required to be composed as that composed partially of the banking system as a whole.
The RIPPLE programs are shown in Figure \ref{ripple_src_webbbs_1} and
Figure \ref{ripple_src_webbbs_2}.

The interface is adopted so that events derived from the HTTP server are
anticipated and the function $bbs\_additem()$ is called for data inputting.
The display start index in the variable $start\_count$ and
the display item count in the variable $list\_count$ are set.
HTML data to be output is always prepared in the variable $output\_html$.
Insertion, deletion, and the exchange of data are performed with
array element operations.

\begin{figure}[tb]
{\footnotesize
$dynamic$ $automatic$ \textbf{struct} $bbsdata\_struct$ \{
\begin{quote}
	string $date =$ ``$2013/01/01$''; \\
	string $name =$ ``''; \\
	string $message =$ ``'';
\end{quote}
\} $[0]$ $bbsdata$; \\
\\
$dynamic$ \textbf{struct} $output\_struct$ \{
\begin{quote}
	string $date$; \\
	string $name$; \\
	string $html$;
\end{quote}
\} $[0]$ $output\_block$; \\
\\
int $start\_count = 0$; \\
int $list\_count = 20$; \\
string $output\_html$; \\
\\
void $bbs\_additem$( string $input\_date$, string $input\_name$,
 string $input\_message$ ) \\
\{
\begin{quote}
	int $i = vsize( bbsdata )$; \\
	$bbsdata[i].date = input\_date[0..3]\;\&$ ``/'' $\&$ \\
	\hspace{0.5cm} $input\_date[5..6]\;\&$ ``/'' $\&\;input\_date[8..9]$; \\
	$bbsdata[i].name = input\_name$; \\
	$bbsdata[i].message = input\_message$; \\
	\textbf{return};
\end{quote}
\}; \\
\hspace{2cm} \vdots \\
}
\caption{Web Bulletin Board System on RIPPLE - 1}
\label{ripple_src_webbbs_1}
\end{figure}

\begin{figure}[tb]
{\footnotesize
$[ \; vsize(output\_block) == (start\_count + list\_count$
\begin{quote}
	$>= vsize(bbsdata) \; ? \; vsize(bbsdata) \; -$
	$start\_count \; : \; start\_count + list\_count); \; ];$ \\
\end{quote}

$[[i] \; output\_block[i].date[0..3] == bbsdata[i+start\_count].date[0..3] \; \&\&$
\begin{quote}
	$output\_block[i].date[4..5] ==$ ``y '' $\&\& $ \\
	$output\_block[i].date[6..7] == bbsdata[i+start\_count].date[5..6] \; \&\&$ \\
	$output\_block[i].date[8..9] ==$ ``m '' $\&\& $ \\
	$output\_block[i].date[10..11] == bbsdata[i+start\_count].date[8..9] \; \&\&$ \\
	$output\_block[i].date[12] ==$ ``d'' $];$ \\
\end{quote}

$[[i] \; output\_block[i].name == strtrunc( bbsdata[i+start\_count].name, 20 ); \; ];$ \\

$[[i] \; output\_block[i].html ==$
\begin{quote}
	``Index:''$ \; \& \; (string)(i+start\_count)$ $\&$ ``$<$BR$>\setminus$n'' $\&$ \\
	``Date:'' $\&$ $output\_block[i].date \; \&$ ``$<$BR$>\setminus$n'' $\&$ \\
	``Name:'' $\& \; output\_block[i].name \; \&$ ``$<$BR$>\setminus$n'' $\&$ \\
	``Message$<$BR$>\setminus$n'' $\&$ $bbsdata[i+start\_count].message \; \&$
	``$<$BR$><$BR$>\setminus$n''; $];$ \\
\end{quote}

$[ \; output\_html ==$ ``$<$HTML$><$BODY$>\setminus$n$\;\;$'' $\&$
\begin{quote}
	$(\mathrm{string})start\_count \; \&$ ``$\;\;$'' $\&$
	$(\mathrm{string})list\_count \; \&$
		``$<$BR$><$BR$>\setminus$n'' $\&$ \\
	$output\_block[*].output\_html \; \&$
		``$</$BODY$></$HTML$>\setminus$n''; $];$ \\
\end{quote}

}
\caption{Web Bulletin Board System on RIPPLE - 2}
\label{ripple_src_webbbs_2}
\end{figure}

Some explanation of
Figure \ref{ripple_src_webbbs_1} and Figure \ref{ripple_src_webbbs_2} is given below.
\begin{itemize}
\item The function $strtrunc()$ truncates characters with overlong strings.
\item \& is the binary operator for the string concatenation operation.
\item \&\& is the binary operator for the logical 'and' operation.
\end{itemize}

\subsection{Conclusion of RIPPLE}

Currently, RIPPLE can process all the sample programs in real-time.
We do not consider speeding up to be as important
as implementing the program language.
RIPPLE-implemented LISP is not used in the implementation of Common LISP,
which has a proven record of accomplishment.
Although this LISP, being a subset of the independently developed Common LISP,
was designed specifically for the RIPPLE implementation,
it has never been considered for managing increasing speeds.
Additionally, due to the limitation in unimplementing non-essential
constraints that can be solved,
the current RIPPLE rewrites these as equivalent operations and executes them.

The web bulletin board executes 12 registration deletions as a batch process in 1 s on a PC running Linux with an Intel Core i5.
Even if cooperating with an Apache web server, a response time of about 1 s is usual, including various overheads.

A big feature of this method is that technology allows it to speed up,
and that we are essentially able to construct a large-scale network system.
Therefore, discussing results in terms of time values is not so significant.

It is important to show that a practical application can be developed by this mechanism.
We believe we have shown it to be capable of achieving a very
practicable processing speed, even though our implementation
did not explicitly consider the speed up of processing.

In this respect, the sample programs showed the descriptive possibility of the application.
In all of the sample programs, the problem was efficiently reformulated without the trial-and-error of the constraint solver,
though only the relations between data are described, thus achieving our purpose.

Unlike the blue algorithm, the method does not require preprocessing beyond the dynamical execution of the constraints when the definition of the variables is changed.

The multi-way constraint computation can handle data processing applications well,
including the batch application described in \cite{spreadsheet}.

\section{Conclusion}

This document has proposed a programming paradigm with non-declarative constraints,
which is a computation model for achieving a balance between a direct response to design methods and program code and effective execution.
The spectrum of the constraint solving algorithm is purple, which is a simplified form of the conventional blue.
This was achieved by focusing the target problem on data operational applications,
typified by a business system, and using the inherent properties.

Correspondence with the design method is maintained even though this model can be modulated by domain decomposition
and removes part of the parallelism of the computing model.
This model can handle the entire system.

\bibliographystyle{acmtrans}
\bibliography{noncstr}

\end{document}